\documentclass[superscriptaddress,preprint,endfloats*,12pt,nofootinbib]{revtex4-1}
\usepackage[utf8]{inputenc}
\usepackage[T1]{fontenc}
\usepackage[american]{babel}
\usepackage{graphicx}
\usepackage{siunitx}

\DeclareSIUnit\torr{Torr}
\DeclareSIUnit\atomic{at.}
\usepackage{booktabs}
\usepackage{tabularx}
\usepackage{miller}

\usepackage{amsmath}
\usepackage[hidelinks]{hyperref}

\usepackage{lineno}

\usepackage{lipsum}
\usepackage{xcolor}

\begin{document}

\title{Experimental quantification of atomically-resolved HAADF-STEM images using EDX}

\author{K.~Pantzas}\email{konstantinos.pantzas@c2n.upsaclay.fr}
\affiliation{Université Paris-Saclay, CNRS, Centre de Nanosciences et de Nanotechnologies - C2N, 91120, Palaiseau, France.}

\author{G.~Patriarche}
\affiliation{Université Paris-Saclay, CNRS, Centre de Nanosciences et de Nanotechnologies - C2N, 91120, Palaiseau, France.}

\begin{abstract}
    Atomically-resolved mappings of the indium composition in InGaN/GaN multi-quantum well structures have been obtained by quantifying the contrast in HAADF-STEM. The quantification procedure presented here does not rely on computation-intensive simulations, but rather uses EDX measurements to calibrate the HAADF-STEM contrast. The histogram of indium compositions obtained from the mapping provides unique insights into the growth of InGaN: the transition from GaN to InGaN and vice versa occurs in discreet increments of composition; each increment corresponds to one monolayer of the interface, indicating that nucleation takes longer than the lateral growth of the step. Strain-state analysis is also performed by applying Peak-Pair Analysis to the positions of the atomic columns identified the quantification of the contrast. The strain mappings yield an estimate of the composition in good agreement with the one obtained from quantified HAADF-STEM, albeit with a lower precision. Possible improvements to increase the precision of the strain mappings are discussed, opening potential pathways for the quantification of arbitrary quaternary alloys at atomic scales.
\end{abstract}

\maketitle
%\linenumbers

\section{Introduction\label{sec:intro}}

Optoelectronic semiconductor devices are complex heterostructures that rely on multiple layers of semiconductor alloys. Tailoring these alloys to meet increasingly demanding designs for modern applications requires a thorough understanding of alloy formation and of the distribution of compositions within individual atomic monolayers and at the interfaces between different materials. A wide variety of characterization techniques can be used to measure the composition of alloys. Few, however, are capable of probing composition with an atomic resolution, such the one achieved in transmission electron microscopy (TEM).

In TEM, the composition of semiconductor alloys can be assessed quantitatively using strain-state analysis \cite{Bartel2008,Gerthsen2003}, cathodoluminescence \cite{Seguin2004, Zagonel:2011cv}, energy-filtered TEM \cite{Lin:2002wb,Davies:2009px}, and energy-dispersive X-ray spectroscopy \cite{Jinschek:2006vn,Pantzas:2011ys}. These techniques either provide high chemical precision or high resolution. Only scanning transmission electron microscopy (STEM) in the high-angle annular dark field (HAADF) mode has been shown to combine high chemical precision with ultimate resolution. Indeed, in HAADF-STEM, the integrated intensity is a function of the atomic number $Z$ of atomic columns scanned by the electron probe \cite{Pennycook:1988xd,Pennycook:2000vx}. This integrated intensity can be simply modeled using Rutherford scattering. In practice, however, several additional factors need to be taken into account and the contrast of a HAADF-STEM micrograph alone can only be interpreted qualitatively  \cite{Lebeau:2008fk,Lebeau:2008uq,Rosenauer2009,Rosenauer2011,Walther:2006kx,Khan:2019ik}. A quantitative interpretation of this contrast requires an external calibration. 
\citet{Rosenauer2011} showed that this calibration can be obtained by comparing normalized experimental intensities to intensities obtained from frozen-lattice simulations. Nonetheless, these simulations tend to be computationally intensive, and adapting them to a variety of material systems can prove prohibitively time consuming.

An alternative means of calibrating the contrast was proposed by the present authors, where the HAADF-STEM contrast is calibrated using  energy-dispersive x-ray diffraction spectroscopy (EDX) measurements. This calibration procedure yielded reliable composition mappings from HAADF-STEM images for two families of semiconductor alloys (InGaN, InAlGaAs), although at magnifications not high enough to resolve individual atomic columns \cite{Pantzas2012,Pantzas2015,Pantzas2016}.

In the present contribution we adapt this procedure to the quantification of atomically resolved HAADF-STEM images. The alloy under investigation is indium gallium nitride (InGaN). Indium gallium nitride is an alloy that is nowadays ubiquitous in our daily life, with applications in light emitting diodes (LEDs), laser diodes and high-power electronics \cite{Nakamura1991,Nakamura1997,Mishra2008,Gil2014,Rajan2013}. Among III-V semiconductors, this alloy holds a particular place, as LEDs based on InGaN/GaN quantum wells exhibit record high internal quantum efficiencies, well beyond \SI{80}{\percent} \cite{Nippert2016}, despite threading dislocation in the \SIrange{1e8}{1e9}{\per\centi\meter\squared} range \cite{Ponce1997}, i.e. four orders of magnitude higher than those typically observed in indium phosphide or gallium arsenide. \citet{Chichibu2006} attributed this robustness to defects to the presence of localized radiative centers, related to the short diffusion length of holes in III-nitrides comparatively to other III-V semiconductors. These localized radiative centers were linked, in theory, to the presence of indium-rich clusters that potentially form quantum dots. Microscopic evidence of their presence has been the subject of intense debates: samples investigated in TEM were shown to form clusters due to damage by the electron beam \cite{Humphreys2007}; indium-rich quantum dots were unambiguously observed in one case in HAADF-STEM \cite{Rosenauer2009}; InGaN wells investigated using atom-probe tomography (APT) \cite{Rigutti2016,Cerezo2007}, on the other hand, showed that InGaN is a random alloy. As a result of these investigations, a wealth of information is available for this alloy, making it a suitable test vehicle for the algorithm presented here.

The paper is structured as follows: the experimental setup is first described. The quantification algorithm is then explained and applied to an atomically--resolved HAADF-STEM image of InGaN quantum wells in GaN barriers. The distribution of indium in the wells and at the interfaces is discussed. Finally, the distribution of indium obtained from the quantification HAADF-STEM contrast is compared against that derived from strain mappings from the same image, as well as EDX measurements.

\section{Experimental setup}

The InGaN sample consists of ten \SI{3}{\nano\meter} thick InGaN wells separated by \SI{8}{\nano\meter} thick GaN barriers. The sample was provided by Novagan. X-ray diffraction measurements confirm that the wells are pseudomorphically accommodated on the GaN substrate and contain \SI{17}{\atomic\percent} indium.

Lamellae for STEM observation were prepared from the sample using Focused Ion Beam (FIB) ion milling and thinning. Prior to FIB ion milling, the sample surface was coated with \SI{50}{\nano\meter} of carbon to protect the surface from the platinum mask deposited used for the ion milling process. Ion milling and thinning were carried out in a FEI SCIOS dual-beam FIB-SEM. Initial etching was performed at \SI{30}{\kilo\electronvolt}, and final polishing was performed at \SI{5}{\kilo\electronvolt}.

All samples were observed in an aberration-corrected FEI TITAN 200 TEM-STEM  operating at \SI{200}{\kilo\electronvolt}. The convergence half-angle of the probe was \SI{17.6}{\milli\radian} and the detection inner and outer half-angles for HAADF-STEM were \SI{69}{\milli\radian} and \SI{200}{\milli\radian}, respectively. The lamella was imaged along the \hkl<11-20> zone axis. All micrographs where 2048 by 2048 pixels. The dwell time was \SI{8}{\micro\second} and the total acquisition time \SI{41}{\second}.

The EDX calibration of the HAADF-STEM contrast is the same as that used in Rereferences~\cite{Pantzas2012,Pantzas2015}. Additional EDX measurements were performed in the Titan microscope featuring the {\color{black} Chemistem system, that uses a Bruker  windowless Super-X four-quadrant detector and has a collection angle of \SI{0.8}{\steradian}}. {\color{black} During EDX acquisition, the sample was also aligned along \hkl<11-20>.} The acquisition time for the linescans was \SI{10}{\minute}, during which no significant drift occurred.

The computer code for the quantification procedure described below has been written for the 64-bit version of Igor Pro 7 (WaveMetrics, Lake Oswego, OR, USA) \footnote{The code can be used on a collaboration basis. A request should be sent by e-mail to the authors.}. In particular, the code uses the Igor Pro Image Processing library for image-processing operations, including particle analysis. {\color{black}Computation-intensive steps, such as extracting the average intensity for each atomic column, have been optimized for multithreaded execution. On a laptop equipped with a quad-core, dual-thread Intel i7 from 2015, i.e. running a total of 8 threads, the most demanding steps take about \SI{15}{\minute} to process.}

\section{Quantification procedure\label{sec:algo}}

The quantification procedure is based on References~\cite{Pantzas2012,Pantzas2015}, where HAADF-STEM and EDX were combined to quantify the composition of InGaN alloys in STEM micrographs acquired at magnifications too low to resolve individual atomic columns. In these references, the composition $x$ of indium in an $\textnormal{In}_x\textnormal{Ga}_{1-x}\textnormal{N}$ alloy was shown to relate to the $\mathcal{Z}$-contrast in the image, $\mathcal{R}$, following~\footnote{For a full derivation see Appendix~\ref{sec:derivation}}:

\begin{equation}
   x = \frac{\mathcal{Z}^{\alpha}_\textnormal{Ga}+\mathcal{Z}^{\alpha}_\textnormal{N}}{\mathcal{Z}^{\alpha}_\textnormal{In}-\mathcal{Z}^{\alpha}_\textnormal{Ga}}\left( \mathcal{R}-1\right).
    \label{eq:composition}
\end{equation}
Here, $\mathcal{Z}_i$ is the atomic number of element $i$ and $\alpha$ a scattering exponent. The contrast $\mathcal{R}$ is computed from the HAADF-STEM image intensity $\mathcal{I}$ following:

\begin{equation}
    \mathcal{R} = \frac{\mathcal{I}-\mathcal{I}_0}{\mathcal{I}_{ref}-\mathcal{I}_0},
    \label{eq:contrastdef}
\end{equation}
where $\mathcal{I}_{0}$ is the HAADF detector background intensity and $\mathcal{I}_{ref}$ is a reference intensity. This reference is evaluated in a region of known composition, for instance in a region of GaN, then linearly extrapolated to the remainder of the image. Correctly defining $\mathcal{I}_{ref}$ is a crucial step, as it also corrects for small variations in the thickness of the lamella. Energy-dispersive X-ray  spectroscopy was used to estimate the scattering exponent $\alpha$, and a value of 1.7 was found to yield the best fit.

\begin{figure}
    \centering
    \includegraphics[width=0.75\textwidth]{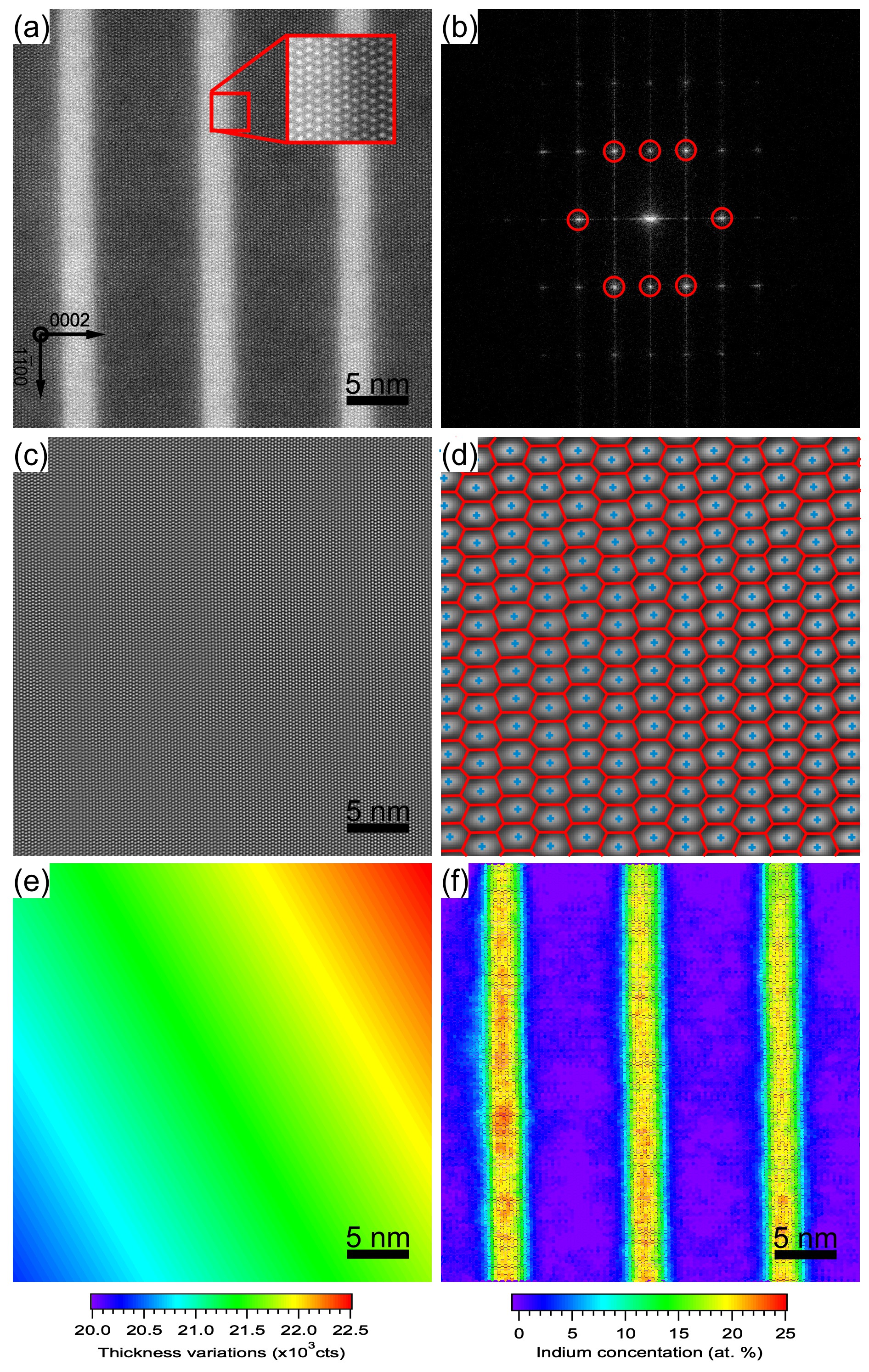}
    \caption{(a) The input to the quantification procedure: an atomically-resolved HAADF-STEM image of the sample. Three InGaN quantum wells in GaN barriers are visible in the image. The inset is focused on the second InGaN well to GaN barrier interface. Dumbbells appear a disks, given that nitrogen has a much lower atomic number than gallium and indium. (b) Power spectral density (PSD) of the micrograph in (a). The first step in the algorithm consists in filtering in the Fourier space. The red circles indicate the crystal reflections that are selected. The remainder of the PSD is filtered out. (c) Bragg-filtered image of the micrograph in (a), obtained from the filtered PSD in (b). Compared to the original micrograph, this image does not contain variations due to composition, and only holds information regarding the position of the disks. This image is used to run a particle analysis and identify the atomic columns present in the image. These positions are in turn used to obtain a partition of the original micrograph using Voronoi tessellation. The image in (d) shows the result of the particle analysis and the tessellation. In the image a magnified portion of the (c) is superimposed with the identified positions of the centers of mass of atomic columns (blue crosses) and the vertices that are the result of the Voronoi tessellation. The identified domains closely resemble the intersection of the Wigner-Seitz cell for InGaN and the \hkl[11-20] planes \cite{Rosenauer2011}. The partition obtained at this step allows one to measure the average HAADF intensity for each atomic column and associate it with the corresponding position. From this point on, the algorithm introduced in Reference~\cite{Pantzas2012} can be applied to quantify the contrast. A crucial step in this quantification procedure is the establishment of a reference intensity in the GaN barriers, i.e. a region of known composition. The intensity measured in GaN is fitted using a plane and extrapolated to the whole image. The result is shown in (e). Using this reference intensity and Equations~\ref{eq:composition} and \ref{eq:contrastdef}, one can now compute the composition mapping in (f).}
    \label{fig:fig1_quanti}
\end{figure}

While at low magnifications it is the intensity of each pixel that is quantified, in atomically-resolved HAADF-STEM micrographs pixels need to be grouped together into domains, each domain encompassing one atomic column. The HAADF-STEM micrograph needs, therefore, to be partitioned prior to applying the quantification procedure described above. Partitioning follows four steps: the original image is first filtered; atomic columns are then identified in the filtered image; the positions of the atomic columns are used to create a partition of the image into domains;  finally, the average intensity over one domain of the partition is attributed to that domain's atomic column. These four steps are formally similar to those described by \citet{Rosenauer2011}, however the specifics of the implementation differ. Figure~\ref{fig:fig1_quanti} illustrates these preprocessing steps. The atomically-resolved HAADF-STEM micrograph in Figure~\ref{fig:fig1_quanti}~(a), shows three InGaN wells, separated by GaN barriers.

The image is filtered in the Fourier space. The power spectral density (PSD) of the image in Figure~\ref{fig:fig1_quanti}~(a) is shown in Figure~\ref{fig:fig1_quanti}~(b). The red circles represent the circular masks that are used for the filtering. They are centered at the most intense crystal frequencies in the PSD. An inverse Fourier transform of the masked PSD yields the Bragg-filtered image, shown in Figure~\ref{fig:fig1_quanti}~(c). Individual atomic columns are identified in the Bragg-filtered image using particle analysis. This step yields a vector of positions $\mathbf{X}$ of all atomic columns in the image. Voronoi tessellation is then applied to $\mathbf{X}$ to obtain a partition of the image into domains centered around the atomic columns. The image in Figure~\ref{fig:fig1_quanti}~(d) shows a magnified portion of the Bragg filtered image of Figure~\ref{fig:fig1_quanti}~(c). The blue crosses represent the centers of the atomic columns, obtained from the particle analysis. The red lines define the domains that result from the Voronoi tessellation. Each red edge is the bisector of a segment defined by two nearest neighbors in $\mathbf{X}$. Therefore, each domain closely resembles the intersection of the Wigner-Seitz cell with the \hkl[11-20] planes \cite{Rosenauer2011}. From this tessellation, the average intensity of each cell is extracted from the original HAADF-STEM image and is associated to the position of the corresponding atomic column.

The reference intensity $\mathcal{I}_{ref}$ is established by  extrapolating the reduced data set of the intensities that correspond to GaN in the barriers to the whole image. The surface plot in Figure~\ref{fig:fig1_quanti}~(d) shows the result of this extrapolation. A gradient from the bottom left to the top right corner is visible in the surface plot. As mentioned above, this gradient is due to variations in the thickness of the lamella. The total variation of thickness accounts for \SI{10}{\percent} of the intensity of GaN, and about \SI{5}{\percent} of the maximum intensity in the image, highlighting how important it is to correct for it. {\color{black} Indeed, without the correction, the thickness gradient would account for an additional variation of \SI{8.5}{\atomic\percent} of indium in the final composition mapping.}

Using the computed reference intensity and by estimating the polarization current following the procedure in Reference~\cite{Pantzas2012}, the normalized contrast $\mathcal{R}$ is computed for each domain of the HAADF-STEM image. The average ratio $\mathcal{R}$ at the core of the InGaN wells is 1.23. Using this value and a scattering exponent of 1.7 yields an average concentration of \SI{17.2}{\atomic\percent} indium, i.e. the nominal of the InGaN wells. Furthermore, the results from frozen lattice simulations in Reference~\cite{Rosenauer2011} also give a ratio of 1.23 for a lamella that is \SI{80}{\nano\meter} thick and contains \SI{17.2}{at.\percent} indium. This indicates that the calibration used in References~\cite{Pantzas2012,Pantzas2015} is valid in the present case also.

\begin{figure}
    \centering
    \includegraphics[width=\textwidth]{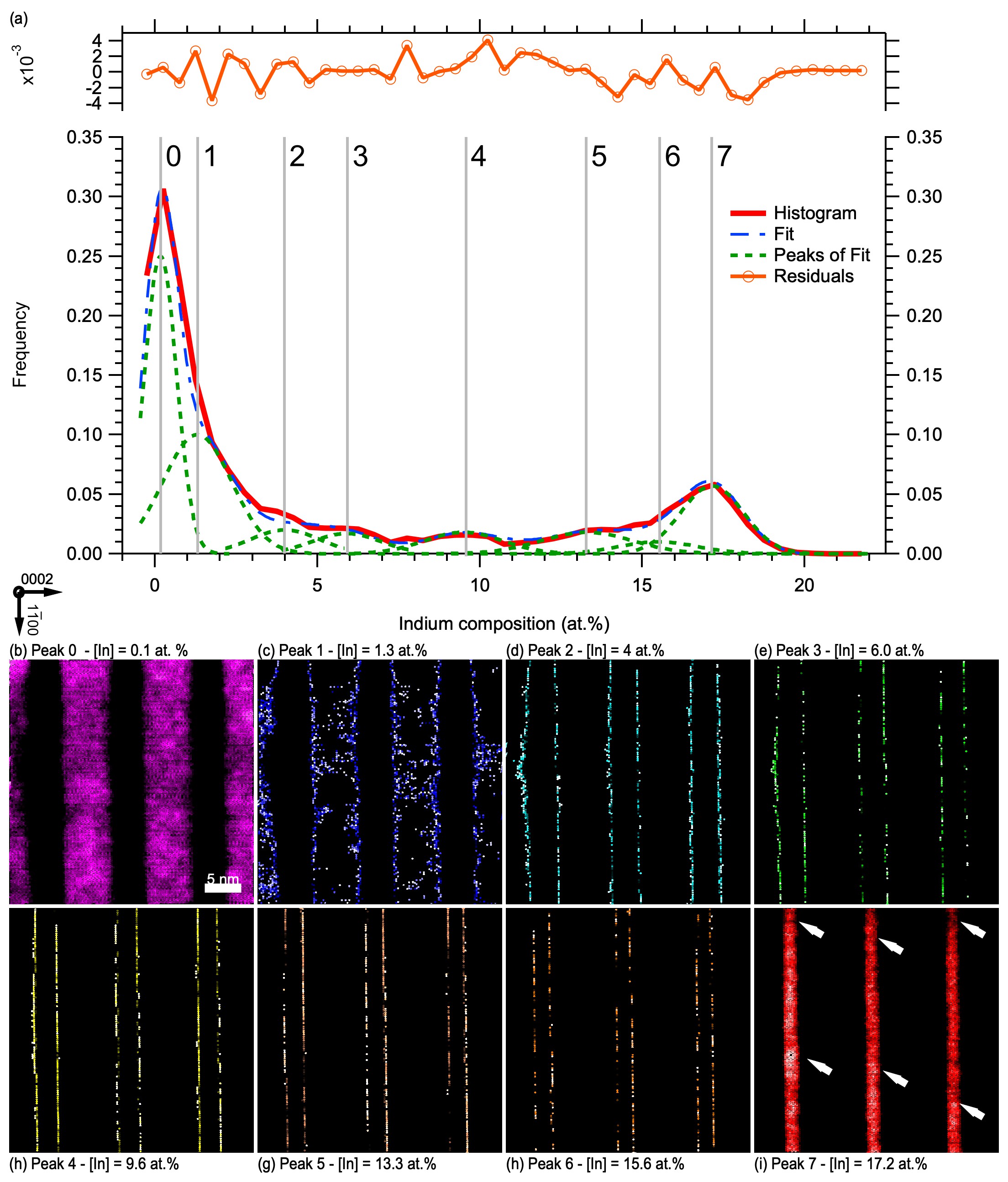}
    \caption{(a) Histogram of the indium compositions from Figure~\ref{fig:fig1_quanti}~(e), represented by the red solid line. The green dashed lines represent peaks identified in the histogram and the blue double-dashed line their sum, fitted to the histogram. {\color{black}Residuals of the fit are also given (orange line with round markers). The residuals show the overall fit is accurate to within \SI{0.5}{\atomic\percent}, a value lower than the precision of the algorithm.} (b-i) are mappings of the composition one standard deviation away from the mean of each peak identified in the histogram. The leftmost peak, centered at \SI{0.1}{\atomic\percent} corresponds to the core of the GaN barriers (b). The standard deviation is \SI{0.8}{\atomic\percent} and corresponds to detector noise. The rightmost peak, centered at \SI{17.2}{\atomic\percent} corresponds to the core of InGaN wells (i). The standard deviation of this peak is \SI{1.7}{\atomic\percent}. It is higher than that of GaN due to random alloy disorder, leading to the appearance of columns that can have composition as far apart \SI{6}{\atomic\percent} (white arrows). Between these two, six additional  peaks can be identified. The mappings (c-h) show that this distinct compositions each correspond to sets of monolayers between the GaN barriers and the InGaN wells. Such monolayers of distinct compositions can only occur during at the interface between GaN and InGaN if the propagation time of an atomic step is significantly lower than the layer's nucleation time. The composition is then fixed to that of the nucleus during nucleation.}
    \label{fig:fig2_hist}
\end{figure}

Using the vector of ratios, Equation~\eqref{eq:composition}, and a scattering exponent of 1.7, the composition mapping presented in Figure~\ref{fig:fig1_quanti}~(e) is finally computed. The mapping reveals the presence of three types of variations of the indium composition: grading along \hkl[0002], at the interfaces between the GaN barriers and the InGaN wells, long-range variations of the indium composition in the InGaN wells along \hkl[1-100], and short-range fluctuations both in InGaN and GaN layers. To better analyze these variations, the histogram of indium compositions is shown in Figure~\ref{fig:fig2_hist}~(a). In addition to the histogram, Figure~\ref{fig:fig2_hist} also presents mappings(b-i) of the spatial distribution of the mean indium concentration for each peak identified in the histogram. The highest peak in the histogram is at \SI{0}{\atomic\percent} and corresponds to the core of the GaN barriers as shown in mapping (b). The standard deviation for GaN is \SI{0.5}{\atomic\percent}, and can be attributed to detector noise \cite{Rosenauer2011,Pantzas2012,Pantzas2015,Pantzas2016}. The noise accounts for the fluctuations observed in the GaN barriers. The second-highest peak is centered at \SI{17.2}{\atomic\percent}, at the opposite end of the histogram. Mapping (i) reveals that the second peak corresponds to the core of the InGaN wells. The standard deviation of this second peak is \SI{1.5}{\atomic\percent}, larger than the standard deviation in GaN. This broadening of the distribution of indium in InGaN is due to random-alloy disorder \cite{Rosenauer2011, Rigutti2016}, although it is lower than the value expected for InGaN that contains \SI{17.2}{\atomic\percent} indium and for a lamella thickness of \SI{80}{\nano\meter}. Indeed, for a random alloy the distribution of compositions is binomial, and the standard deviation, $\sigma_\textnormal{alloy}$, is \SI{2.4}{\atomic\percent}. Taking into account broadening due to detector noise, measured in GaN, the total standard deviation becomes \SI{2.5}{\atomic\percent}. \citet{Rosenauer2011} attributed this discrepancy to averaging of the intensity from contribution of neighboring. An alternate interpretation, however, is possible. Indeed, given the  thickness of the lamella, a single column can cross several atomic steps. The composition in each step is expected to vary slightly due to random alloy disorder. Therefore, the intensity measured for each column will represent an average over $N$ steps, and the standard deviation is reduced by a factor $\sqrt{1/N}$. Using the values cited above, averaging would need to occur over 2.6 atomic steps, a reasonable estimate of atomic steps crossing the atomic column under observation in an \SI{80}{\nano\meter} thick lamella. Given this standard deviation, what may appear to be indium clusters fall well within compositions expected for an InGaN alloy with an average indium content of \SI{17.2}{\atomic\percent}. In fact, atomic columns with compositions as far apart as \SI{6}{\atomic\percent} are likely to be observed, as they lie within the $\left[\mu_\textnormal{In}+2\sigma_\textnormal{In},\mu_\textnormal{In}-2\sigma_\textnormal{In}\right]$ interval. Examples of such clusters can be seen in the areas pointed out by white arrows in mapping (i) of Figure~\ref{fig:fig2_hist}. The clusters in the same well are far apart enough to create localized emission centers, given the low hole diffusion lengths typically observed in InGaN alloys \cite{Chichibu2006,Callsen:2019hf}, without any other phase separation or clustering mechanism to have to be taken into account. 

Between the two aforementioned peaks of GaN and InGaN, the histogram in Figure~\ref{fig:fig2_hist}~(a) is not typical of homogeneously graded interfaces between InGaN and GaN. Instead of continuous tail, the histogram displays additional peaks at distinct indium compositions. To obtain a good fit to the histogram, a total of six more peaks needs to be added to the sum of Gaussian curves used to fit the peaks of the cores of the GaN barriers and the InGaN wells. {\color{black} Gaussian curves are used as the distribution of compositions in InGaN is expected to follow a normal distribution. As stated above, InGaN is a random alloy and the composition in a single atomic column follows by a binomial distribution. By virtue of the central-limit theorem, this binomial distribution is well approximated by a normal distribution. Given that each frequency maps back to several monolayers and that each monolayer contains more than one hundred atomic columns, the conditions for the central-limit theorem apply here.} Mappings (c) through (h) reveal that each peak corresponds to monolayers of distinct composition in the transition from GaN to InGaN and, symmetrically, from InGaN to GaN. These monolayers are not necessarily contiguous, but on average extend laterally to about \SI{25}{\nano\meter}, a value that is consistent with the typical step size in the layer. This indicates that the nucleation time is significantly shorter than the propagation time of a step. {\color{black}The composition of each layer in the transition is fixed at the time of its nucleation relative to the concentration of the indium precursor in the vapor phase. This concentration varies briefly in the time it takes to commute the indium source between the growth and vent manifolds in the MOCVD reactor.}

\section{Comparison with alternate methods for measuring the composition\label{sec:comparison}}

The composition of alloys in an atomically resolved HAADF-STEM image can also be deduced from strain mappings computed from the image, albeit with a lower precision. Several algorithms have been developed over the years to compute such strain mappings, including geometric-phase analysis (GPA) \cite{Hytch1998}, peak-pairs analysis (PPA) \cite{Galindo2007}, and template matching (TeMa) \cite{Zuo2014}. Of the three, the PPA algorithm can be straightforwardly implemented in the present case: in PPA, strain is computed from the real-space positions of atomic columns. These positions have already been retrieved in the process of quantifying the HAADF-STEM contrast. The algorithm has been modified slightly from the one in Reference~\cite{Galindo2007}, in an attempt to improve its precision. In reference, the displacement is calculated between a given point and it's nearest neighbor along one of the two base vectors that have been chosen. Here, the displacement is instead computed as half the sum of the displacements from a given point to it's nearest neighbor along a base vector and from the same point to it's nearest neighbor in the direction opposite of the chosen base vector. This second definition is tantamount to using central differences to compute the displacement, which are well known to yield more precise results than forward or backward differences for the same number of grid points\footnote{For further information please refer to  Appendix~\ref{sec:ppa}}.

\begin{figure}
    \centering
    \includegraphics[width=\textwidth]{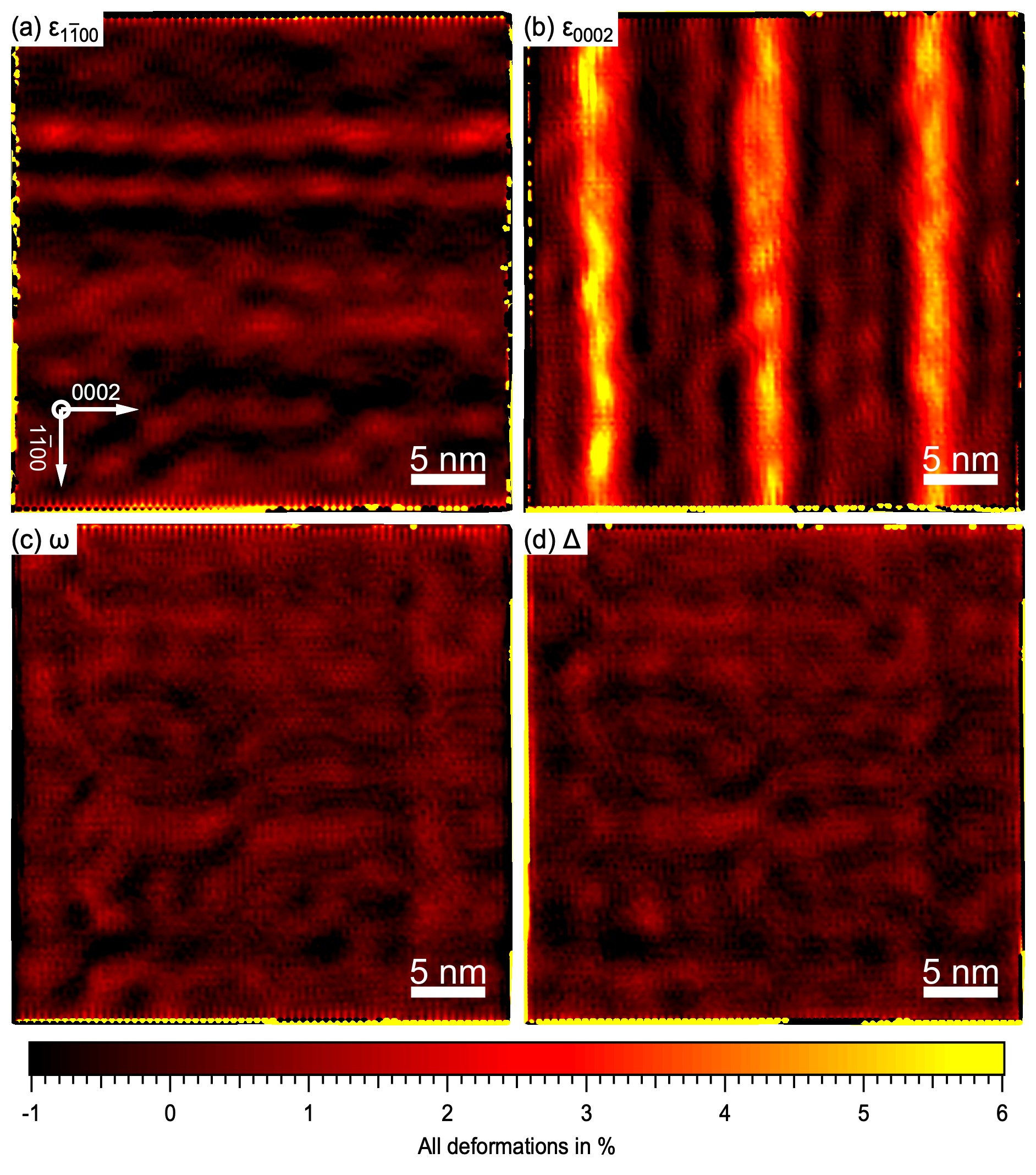}
    \caption{Deformation mappings, computed from the HAADF-STEM in Figure~\ref{fig:fig1_quanti}~(a), using the modified PPA algorithm described in the text. The mappings show that there is strain in InGaN only along \hkl0002, i.e. the growth direction. The average strain in the core of the InGaN wells is \SI{5.2}{\percent}, a value in good agreement with an indium composition of \SI{17.2}{\atomic\percent} in InGaN layers that are pseudomorphically accommodated on GaN. The error, measured in the GaN barrier, is \SI{0.3}{\percent}. It is mainly due to fluctuations that are the result of a small residual magnetic field in the environment of the microscope. Such fluctuations can, in certain case be corrected for, increasing the precision of the strain-state analysis, and, as a result that of composition estimates derived from it \cite{Sanchez2005}.}
    \label{fig:fig3_def_maps}
\end{figure}

Furthermore, it is well established that in HAADF-STEM there is a small amount of distortion that is related to sample drift during acquisition \cite{Zuo2014}. To obtain precise results, this drift has to be taken measured and corrected for. The correction is based on the assumption that overall drift is slow and does not change in direction during the acquisition. The drift velocity is evaluated in regions of GaN, where no strain is expected. This velocity is then used to correct the positions in vector $\mathbf{X}$ before applying the modified PPA algorithm. The end result are the four mappings shown in Figure~\ref{fig:fig3_def_maps}, that were obtained from the HAADF-STEM micrograph in Figure~\ref{fig:fig1_quanti}~(a). In Figure~\ref{fig:fig3_def_maps}, strain mapping (a) represents the in-plane strain, {\color{black} $\varepsilon_{1\bar{1}00}$}, mapping (b) the strain along the growth direction, $\varepsilon_{0002}$,  mapping (c) the rotation and mapping (d) the shear strain. Of the four, only mapping (b) shows the presence of strain, i.e. the InGaN wells are pseudomorphically accommodated on GaN. The average strain in the core of the wells is \SI{5.3}{\percent}, a value in good agreement with that expected for an InGaN alloy with an average indium composition of \SI{17.2}{\percent}. The precision of the mappings, measured as the standard deviation in the GaN areas is \SI{0.4}{\percent}. 

\begin{figure}
    \centering
    \includegraphics[width = \textwidth]{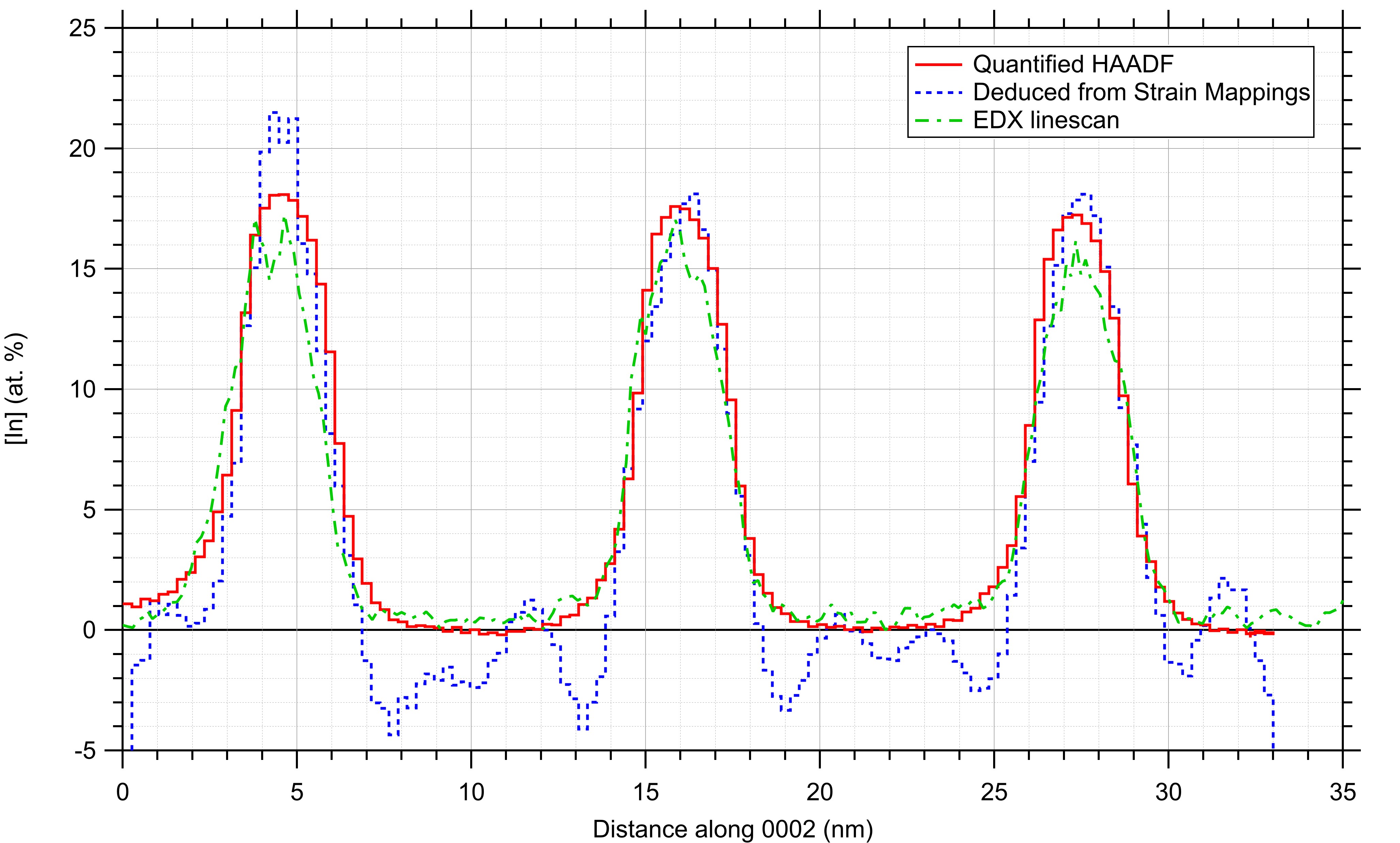}
    \caption{Plot comparing the indium composition per \hkl[1-100] monolayer, measured using three different techniques: the red solid line corresponds to quantified HAADF-STEM, the blue dashed curve to the composition deduced from strain-state analysis, and the green dot-dashed curve an EDX linescan. Of the three, quantified HAADF-STEM is shown to provide the most precise measurement, with a measurement error in GaN of less than \SI{0.1}{\atomic\percent}. Although strain-state analysis yields a composition in good agreement with quantified HAADF-STEM, fluctuations are higher as a result of the magnetic environment of the microscope's room. Finally, the EDX linescan is shown to provide a good estimate as well, with a residual error of about \SI{0.5}{\atomic\percent}. The wells do not show significant broadening.}
    \label{fig:fig4_comparison}
\end{figure}

Using the strain mappings of Figure~\ref{fig:fig3_def_maps} and the method for determining the composition from strain described in Reference~\cite{Schuster1999}, a second estimate of the composition is obtained from the same image. The composition measured in this manner is compared against the one obtained by quantifying the contrast in the HAADF-STEM micrograph in the plot of Figure~\ref{fig:fig4_comparison}. The plot graphs the average indium concentration per monolayer, i.e. the average over columns that belong to the same \hkl(0002) plane. The red solid curve represents the composition determined by quantifying the HAADF-STEM contrast and the blue dashed curve the composition determined from the strain-state analysis. The two curves are in remarkable agreement. Nevertheless, the composition from HAADF-STEM is shown to be more precise: fluctuations in GaN are less than \SI{0.1}{\atomic\percent}, as each monolayer groups more than 100 atomic columns. In the case of the strain-state analysis, though GaN on average is close to \SI{0}{\atomic\percent}, the standard deviation in GaN from strain analysis is closer to \SI{1.5}{\atomic\percent}. Indeed, quasi-periodic oscillations are visible in GaN. Such fluctuations are the result of a small residual magnetic field present in the microscope's room, that affects the electron-beam during acquisition, and that despite the presence of a magnetic compensation loop. Having such a residual is unavoidable, but there have been recent successful attempts to correct for it \cite{Sanchez2005,Sanchez2006}. In the present case, however, applying the method described in these references, was not possible, as fluctuations appear across both axes of the image. {\color{black} A few workarounds have been proposed in the literature. One approach consists in acquiring a single image at \SI{45}{\degree} angle to the scanning direction, thereby allowing one to fit and correct for these fluctuations in both directions in a fashion similar to the one proposed in Reference~\cite{Sanchez2005}. A second approach consists in acquiring image pairs with orthogonal scan directions, allowing a better estimate of distortion \cite{Ophus2016}. Finally, revolving STEM, where a series of images acquired at several scanning angles was proposed to correct drift and distortions without prior knowledge \cite{Sang2014}. In these methods increased precision comes at the cost of increased complexity of the acquisition. Alternatively, newer acquisition modes implemented in state of the art microscope where the microscope automatically acquires a series of rapid images, corrects for drift between them and sums them are now becoming available and are a promising way to decrease the residual errors in strain analysis.} Such improvements would increase the precision of composition estimates based on strain-state analysis. The threshold for the precision of strain-state analysis to yield composition estimates comparable to that of quantified HAADF-STEM is a few tens to a hundred ppm, i.e. at least an order of magnitude better than the current precision. Reaching such precision is highly desirable, as it would allow one to quantify arbitrary quaternary alloys, where strain-state analysis or the intensity alone would not suffice to access both compositions of the alloy.

Finally, the third curve in the plot of Figure~\ref{fig:fig4_comparison} graphs the results of an EDX linescan across the same area. The measured composition is in excellent agreement with that measured in quantified HAADF-STEM, although the noise floor is higher: the residual average in GaN is \SI{0.4}{\atomic\percent} with a standard deviation of \SI{0.2}{\atomic\percent}. No broadening of the wells is observed, indicating that at this magnification convolution with the electron beam is minimal. This is an encouraging result that shows that it is possible to use EDX to precisely measure the composition even in thin wells. It does not, however, provide quantitative information on the composition on the level of the atomic column, making it as of yet impossible to evaluate the distribution of compositions in an image and the insights that distribution provide regarding the growth mechanism of the layers in the image.

\section{Conclusions}

In the present contribution, atomically-resolved, quantitative mappings of the indium composition in InGaN/GaN multi-quantum well structures have been computed from HAADF-STEM micrographs. The quantification procedure does not rely on simulations, but rather a calibration through comparison with EDX measurements, a method that can be straightforwardly applied to a variety of semiconductor and other alloys. The resulting mapping contains a wealth of information regarding the distribution of indium in the structure, providing a wealth of insights as to its growth mechanism. The histogram of compositions in the mapping reveals that the transition from GaN to InGaN and vice occurs over monolayers of distinct compositions, a fact that shows that the composition is fixed at the nucleation, i.e. that the nucleation time of a layer is significantly higher than the subsequent lateral growth. Strain-state analysis was also performed by applying a modified version of the PPA algorithm to the position vector of identified atomic columns, obtained during the quantification. The composition deduced from the strain mappings is in good agreement with that from quantified HAADF-STEM. The precision is, however, at the moment lower. Increasing the precision of the strain-state analysis would open the path for the precise quantification at the atomic scale of arbitrary quaternary alloys.

\section*{Acknowledgments}
The authors would like to thankfully acknowledge funding from the CNRS Renatech network, the ANR Labex TEMPOS (ANR-10-EQPX-0050), and the ANR project LIGNEDEMIR (ANR-18-CE09-0035).

\appendix

\section{Relationship between HAADF contrast and composition\label{sec:derivation}}

The HAADF-STEM annular detector detects electrons scattered to high angles through Rutherford scattering. The detected  intensity $\mathcal{I}$ can be expressed as follows:

\begin{equation}
    \mathcal{I}-\mathcal{I}_0 = K d\sum_{i}x_i\mathcal{Z}_i^{\alpha},
    \label{eq:haadfInt}
\end{equation}

where $\mathcal{I}_0$ is the background intensity, and depends on the brightness settings used to acquire the HAADF-STEM image. $K$ is a constant, $d$ the thickness of the lamella under observation, $\left(x_i,\mathcal{Z}_i\right)$ the concentration and atomic number of scattering element $i$ in the lamella, and $\alpha$ the scattering exponent. Following Equation~\eqref{eq:haadfInt}, the intensity of $\textnormal{In}_x\textnormal{Ga}_{1-x}\textnormal{N}$ alloy is:

\begin{equation}
    \mathcal{I}-\mathcal{I}_0 = K  d \left(x\mathcal{Z}^{\alpha}_\textnormal{In}+\left(1-x\right)\mathcal{Z}^{\alpha}_\textnormal{Ga}+\mathcal{Z}^{\alpha}_\textnormal{N}\right).
    \label{eq:intInGaN}
\end{equation}

the contrast $\mathcal{R}$ is defined as the ratio of the intensity in Equation~\eqref{eq:haadfInt} to a reference intensity $\mathbf{I}_{ref}$:

\begin{equation}
    \mathcal{R} = \frac{\mathcal{I}-\mathcal{I}_0}{\mathcal{I}_{ref}-\mathcal{I}_0}.
    \label{eq:contrastdef_ap}
\end{equation}

If the reference intensity is taken in GaN, and linearly extrapolated to the whole image, one can show that $\mathcal{R}$ is linked to the composition $x$ through:

\begin{equation}
    \mathcal{R} = \frac{\mathcal{Z}^{\alpha}_\textnormal{In}-\mathcal{Z}^{\alpha}_\textnormal{Ga}}{\mathcal{Z}^{\alpha}_\textnormal{Ga}+\mathcal{Z}^{\alpha}_\textnormal{N}}x+1,
    \label{eq:contrast}
\end{equation}

or, equivalently:

\begin{equation}
   x = \frac{\mathcal{Z}^{\alpha}_\textnormal{Ga}+\mathcal{Z}^{\alpha}_\textnormal{N}}{\mathcal{Z}^{\alpha}_\textnormal{In}-\mathcal{Z}^{\alpha}_\textnormal{Ga}}\left( \mathcal{R}-1\right).
    \label{eq:composition_ap}
\end{equation}

Using EDX and this equation, one can estimate the scattering exponent $\alpha$. In the case of InGaN, we have shown that a value of 1.7 yields the best fit \cite{Pantzas2012,Pantzas2015}.

\section{Modified PPA Algorithm\label{sec:ppa}}

In Peak-Pair Analysis (PPA), the strain fields $\varepsilon_{xx}$, $\varepsilon_{xy}$, $\varepsilon_{yy}$, and $\varepsilon_{yx}$ are computed from the displacements $\left(\mathbf{U},\mathbf{V}\right)$ between nearest neighbors. Four nearest neighbors to a given position $\left(x,y\right)$ in the vector of atomic-column positions $\mathbf{X}$ are identified using two non-colinear base vectors $\left(\vec{\mathbf{a}},\vec{\mathbf{b}}\right)$. In the following, the index pair $\left(n,m\right)$ is used to denote the position at which the displacements $\left(\mathbf{U},\mathbf{V}\right)$ are computed. The nearest neighbor in the direction of base vector $\vec{\mathbf{a}}$ is denoted using the index pair $\left(n+1,m\right)$ and the nearest neighbor in the direction of base vector $\vec{\mathbf{b}}$ the index pair $\left(n,m+1\right)$. Using this notation, \citet{Galindo2007} define the deformation $\mathbf{U}$ with respect to base vector $\vec{\mathbf{a}}$ at position $\mathbf{X}_{n,m}$ as:
 
 \begin{align}
    u_x &= x_{n+1,m}-x_{n,m}-a_x \label{eq:ux_fd}\\
    u_y &= y_{n+1,m}-y_{n,m}-a_y.\label{eq:uy_fd}
\end{align}
 
 Similarly, deformation $\mathbf{V}$ with respect to base vector $\vec{\mathbf{b}}$ at position $\mathbf{X}_{n,m}$ is defined as:
 
 \begin{align}
    v_x &= x_{n,m+1}-x_{n,m}-b_x \label{eq:vx_fd}\\
    v_y &= y_{n,m+1}-y_{n,m}-b_y.\label{eq:vy_fd}
\end{align}
 
 Given, however, that the nearest neighbors $\left(n-1,m\right)$, and $\left(n,m-1\right)$, along the directions of vectors $-\vec{\mathbf{a}}$ and $-\vec{\mathbf{b}}$, respectively, have also been identified, the displacements $\left(\mathbf{U},\mathbf{V}\right)$ can also be defined as:

\begin{align}
    u_x &= x_{n,m}-x_{n-1,m}+a_x \label{eq:ux_bd}\\
    u_y &= y_{n,m}-y_{n-1,m}+a_y\label{eq:uy_bd}\\
    \intertext{and}
    v_x &= x_{n,m}-x_{n,m-1}-b_x \label{eq:vx_bd}\\
    v_y &= y_{n,m}-y_{n,m-1}-b_y.\label{eq:vy_bd}
\end{align}

To increase the precision of the strain mappings, the deformation in the present contribution is taken as the average between Equations~\eqref{eq:ux_fd}-\eqref{eq:vy_fd} and Equations~\eqref{eq:ux_bd}-\eqref{eq:vy_bd}, i.e. $\left(\mathbf{U},\mathbf{V}\right)$ are expressed as follows:

\begin{align}
    u_x &= \frac{1}{2}\left(x_{n+1,m}-x_{n-1,m}\right)+a_x \label{eq:ux_cd}\\
    u_y &= \frac{1}{2}\left(y_{n+1,m}-y_{n-1,m}\right)+a_y\label{eq:uy_cd}
    \intertext{and}
    v_x &= \frac{1}{2}\left(x_{n,m+1}-x_{n,m-1}\right)-b_x \label{eq:vx_cd}\\
    v_y &= \frac{1}{2}\left(y_{n,m+1}-y_{n,m-1}\right)-b_y.\label{eq:vy_cd}
\end{align}

Using Equations~\eqref{eq:ux_cd}-\eqref{eq:vy_cd} instead of Equations~\eqref{eq:ux_fd}-\eqref{eq:vy_fd} is equivalent to using central finite differences instead of forward finite differences when one computes the strain fields. Indeed, the strain fields are defined as the partial derivatives of the displacement versus the position:

\begin{equation}
    \varepsilon_{xx} = \frac{\partial \mathbf{U}}{\partial x} , \quad \varepsilon_{xy} = \frac{\partial \mathbf{U}}{\partial y} , \quad \varepsilon_{yy} = \frac{\partial \mathbf{V}}{\partial x} , \quad \varepsilon_{yx} = \frac{\partial \mathbf{V}}{\partial y},
    \label{eq:straindef}
\end{equation}

and, in the original PPA paper, the strain fields are computed by solving the following set of linear equations:
\begin{equation}
\begin{alignedat}{2}
    a_x \varepsilon_{xx} & + a_y \varepsilon_{xy} && = u_x \\
    a_y \varepsilon_{yy} & + a_x \varepsilon_{yx} && = u_y \\
    b_x \varepsilon_{xx} & + b_y \varepsilon_{xy} && = v_x \\
    b_y \varepsilon_{yy} & + b_x \varepsilon_{yx} && = v_y.\label{eq:epssyst}
\end{alignedat}
\end{equation}
As a result, Equations~\eqref{eq:ux_fd}-\eqref{eq:vy_fd} and  \eqref{eq:epssyst} compute the strain fields from the positions using forward differences, whereas  Equations~\eqref{eq:ux_cd}-\eqref{eq:vy_cd} and  \eqref{eq:epssyst} compute the strain fields from the positions using central differences. Forward differences are $O(h)$ accurate, whereas central differences are $O(h^2)$ \cite{Press:2007kh}, hence, the second set of equations yields more accurate values for the strain fields from the same vector of positions.

\section{Corrections for drift in the STEM image\label{sec:drift}}

In STEM, images are acquired by scanning the probe over a region of interest in the sample. Even in the most stable environments, small sample drift still occurs. This results in slight distortions in the image that need to be corrected for when one computes strain from a given HAADF-STEM image. Using the same notation as in Section~\ref{sec:ppa} and assuming small drift in an arbitrary direction, the position of an atomic column in the image can be expressed as follows:

\begin{equation}
\begin{alignedat}{2}
    x_{n,m} &= x_{0,0} + n a_x +m b_x\\
    %\phantom{x_{n,m}} & \phantom{=} +\left(x_{n,m}-x_{0,0}\right) c_x \delta t_d\\
    %\phantom{x_{n,m}} & \phantom{=} +\left(y_{n,m}-y_{0,0}\right) c_y \delta t_l\\
    \phantom{x_{n,m}} & \phantom{=} + u_x(x_{n,m},y_{n,m}) \\
    \phantom{x_{n,m}} & \phantom{=} + v_x(x_{n,m},y_{n,m}) \\
    \phantom{x_{n,m}} & \phantom{=} + \delta x_d
\end{alignedat}
\end{equation}
and
\begin{equation}
\begin{alignedat}{2}
    y_{n,m} &= y_{0,0} + n a_y +m b_y\\
    %\phantom{x_{n,m}} & \phantom{=} +\left(x_{n,m}-x_{0,0}\right) c_x \delta t_d\\
    %\phantom{x_{n,m}} & \phantom{=} +\left(y_{n,m}-y_{0,0}\right) c_y \delta t_l\\
    \phantom{x_{n,m}} & \phantom{=} + u_y(x_{n,m},y_{n,m}) \\
    \phantom{x_{n,m}} & \phantom{=} + v_y(x_{n,m},y_{n,m}) \\
    \phantom{x_{n,m}} & \phantom{=} + \delta y_d.
\end{alignedat}
\end{equation}

Here, $\delta x_d$ and $\delta y_d$ represent small drift in the $x$ and $y$ directions of the image, respectively. Assuming the drift velocity $\vec{\mathbf{c}}$ to be constant during image acquisition, these $\delta x_d$ and $\delta y_d$ can be expressed as a function of the pixel dwell time $\delta t_d$ and the line acquisition time $\delta t_l$:

\begin{align}
\delta x_d = c_x\left(\lfloor x_{n,m}\rfloor \delta t_d + \lfloor y_{n,m}\rfloor \delta t_l\right) \label{eq:driftx}
\intertext{and}    
\delta y_d = c_y\left(\lfloor x_{n,m}\rfloor \delta t_d + \lfloor y_{n,m}\rfloor \delta t_l\right). \label{eq:drifty}
\end{align}

Here $\lfloor.\rfloor$ denotes the floor function. Equations~\eqref{eq:driftx} and \eqref{eq:drifty} express the drift as the drift velocity multiplied by the time it took the beam to reach the pixel where the atomic columns was identified. Based on these expressions, the displacements $\left(\mathbf{U},\mathbf{V}\right)$ become:

%\begin{equation}
\begin{alignat}{2}
    u_x &= \frac{1}{2}\left(x_{n+1,m}-x_{n-1,m}\right)+a_x\nonumber\\
    \phantom{u_x} &\phantom{=}+\frac{1}{2} c_x  \delta t_d\lfloor x_{n+1,m}-x_{n+1,m}\rfloor \label{eq:ux_cd_corr}\\
    \phantom{u_x} &\phantom{=}+\frac{1}{2} c_x  \delta t_l\lfloor y_{n+1,m}-y_{n-1,m}\rfloor\nonumber\\
    u_y &= \frac{1}{2}\left(y_{n+1,m}-y_{n-1,m}\right)+a_y\nonumber\\
    \phantom{u_x} &\phantom{=}+\frac{1}{2} c_y  \delta t_d\lfloor x_{n+1,m}-x_{n+1,m}\rfloor \label{eq:uy_cd_corr}\\
    \phantom{u_x} &\phantom{=}+\frac{1}{2} c_y  \delta t_l\lfloor y_{n+1,m}-y_{n-1,m}\rfloor\nonumber\\
    \intertext{and}
    v_x &= \frac{1}{2}\left(x_{n,m+1}-x_{n,m-1}\right)-b_x\nonumber\\
    \phantom{u_x} &\phantom{=}+\frac{1}{2} c_x  \delta t_d\lfloor x_{n+1,m}-x_{n+1,m}\rfloor \label{eq:vx_cd_corr}\\
    \phantom{u_x} &\phantom{=}+\frac{1}{2} c_x  \delta t_l\lfloor y_{n+1,m}-y_{n-1,m}\rfloor \nonumber\\
    v_y &= \frac{1}{2}\left(y_{n,m+1}-y_{n,m-1}\right)-b_y\nonumber\\
    \phantom{u_x} &\phantom{=}+\frac{1}{2} c_y  \delta t_d\lfloor x_{n+1,m}-x_{n+1,m}\rfloor \label{eq:vy_cd_corr}\\
    \phantom{u_x} &\phantom{=}+\frac{1}{2} c_y  \delta t_l\lfloor y_{n+1,m}-y_{n-1,m}\rfloor.\nonumber
\end{alignat}
%\end{equation}

In an unstrained reference region of the HAADF-STEM image there are no displacements except for the drift terms, and Equations~\eqref{eq:ux_cd_corr}-\eqref{eq:vy_cd_corr} are simplified to:

\begin{align}
    u_x^{ref} &= \frac{1}{2} c_x  \delta t_d\lfloor x_{n+1,m}-x_{n+1,m}\rfloor + \frac{1}{2} c_x  \delta t_l\lfloor y_{n+1,m}-y_{n-1,m}\rfloor \label{eq:ux_cd_ref}\\ 
    u_y^{ref} &= \frac{1}{2} c_y  \delta t_d\lfloor x_{n+1,m}-x_{n+1,m}\rfloor + \frac{1}{2} c_y  \delta t_l\lfloor y_{n+1,m}-y_{n-1,m}\rfloor\label{eq:uy_cd_ref}\\
    \intertext{and}
    v_x^{ref} &= \frac{1}{2} c_x  \delta t_d\lfloor x_{n+1,m}-x_{n+1,m}\rfloor + \frac{1}{2} c_x  \delta t_l\lfloor y_{n+1,m}-y_{n-1,m}\rfloor \label{eq:vx_cd_ref}\\
    v_y^{ref} &= \frac{1}{2} c_y  \delta t_d\lfloor x_{n+1,m}-x_{n+1,m}\rfloor + \frac{1}{2} c_y  \delta t_l\lfloor y_{n+1,m}-y_{n-1,m}\rfloor.\label{eq:vy_cd_ref}
\end{align}

One can obtain least-squares estimator of the drift velocity components, $c_x$ and $c_y$, from Equations~\eqref{eq:ux_cd_ref}-\eqref{eq:vy_cd_ref}, and  use these values to correct displacements and strain fields for sample drift.

\bibliographystyle{unsrtnat}
\bibliography{QAS_UM_biblio}

\end{document}